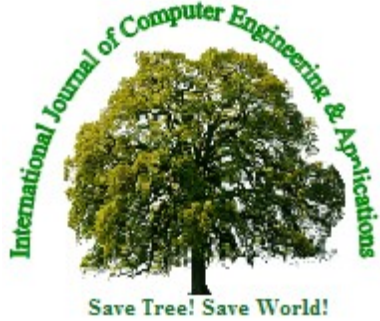

# A NOVEL ARCHITECTURE FOR QUESTION CLASSIFICATION BASED INDEXING SCHEME FOR EFFICIENT QUESTION ANSWERING

Renu Mudgal [1], Rosy Madaan [2], A.K.Sharma[3], Ashutosh Dixit[4]
*1, 2 Department of Computer Science & Engineering*
*Echelon Institute of Technology*
*Faridabad, India*
*3, 4. Department of Computer Engineering*
*Y.M.C.A. University of Science &Technology*
*Faridabad, India*

## ABSTRACT

Question answering system can be seen as the next step in information retrieval, allowing users to pose question in natural language and receive compact answers. For the Question answering system to be successful, research has shown that the correct classification of question with respect to the expected answer type is requisite. We propose a novel architecture for question classification and searching in the index, maintained on the basis of expected answer types, for efficient question answering. The system uses the criteria for "Answer Relevance Score" for finding the relevance of each answer returned by the system. On analysis of the proposed system, it has been found that the system has shown promising results than the existing systems based on question classification.

## [I] INTRODUCTION

Question Answering (QA) system [2,8] is the next step in information retrieval [4,7]. As compare to search engine, QA system retrieves answer to user's question rather than retrieving the whole document. In QA system, the main issue for the researchers is to provide accurate answer from huge collection of information on the Web. The QA system as a whole to be successful, the correct classification of the question is requisite. Question Classification [3,9,10,11] is technique used to extract useful information from the question by identifying its class. Then, to provide user with relevant set of answers, the appropriate answer type needs to be identified on the basis of user's expectation. If the user asks *"Who is the First Prime Minister of India*?", the user expects *"Pandit Jawaharlal Nehru" as* the answer which is the name of a person. For this, the question class *"Who"* is mapped to the expected answer type i.e. *Person*.

Blogs Pages are the richest source of information where people express their opinion on various topics and situations. Blog written by one may be shared by others. Blogs are search friendly and attract more traffic for its fresh and dynamic contents. Blogs have built-in RSS (Really Simple Syndication) Feed that syndicates recent posts and delivers them to various blog search engines. When a reader subscribes to RSS Feed, the recent Blog post is delivered to his mailbox automatically. Most of the Blog pages are updated frequently and the Blog post are sorted in descending order of the date on which they have been written. Blog posts are written by experienced person called Blogger [12] on various topics. So, the content on the Blog pages is likely to be related to the topic on which the Blog is written. This arise an issue of crawling of the Blog content. Here, the Blog oriented crawler come into picture. The crawler used for crawling Blog pages uses features of a Blog page that differ from a general Web page.

Summarization [1] is a technique used to extract sentences from a text document that best represents its content. The summary generated given an idea of what the document is about and what are its contents. Summarization is very useful for filtering out non-relevant content from a text document. It helps a user in deciding whether the document is of his use or not. Two techniques are used for summarization; content based and context based summarization. The content-based summarization utilizes textual content of the web document while context-based makes use of the hypertext structure of the web.

This paper has been organized in following sections: section 2 describes the current research that has been carried out in this area; section 3 discusses the proposed work, section 4 gives the experimental evaluation, section 5 illustrates the discussions and the last section concludes the proposed work.

## [II] RELATED WORK

Huang et al. in [9] discussed a classification using Head Words and their Hypernyms where two models of classifier are used namely Support vector machine and maximum entropy model. Support Vector Machine is a useful technique for data classification. It uses kernel function for problem solution. Four basic kernel functions are linear, polynomial, radial basis function, and sigmoid. Maximum Entropy Model is also known as log-linear and exponential learning models which provides a general purpose machine learning technique for classification which can be successfully applied to natural language processing including part of speech tagging, named entity recognition etc. Here, each question is represented as a bag of features like feature sets, namely question wh-word, head word, WordNet semantic features for head word, word grams, and word shape feature.

Bu et al. in [11] proposed a function-based question classification technique in which question classifier based on MLN (Markov Logic Network) is included. A function-based question classification category tailored to general question answer. The category contains six types, namely Fact, List, Reason, Solution, Definition and Navigation. Each question is split into functional words and content words. The strict pattern from functional words and soft patterns from content words is generated. The matching degree is either 0 or 1 for strict pattern. Finally, markov logic network (MLN) is used to combine and evaluate all the patterns. The function-based taxonomy tailored to general question answering system by using two principles. First, questions can be distributed into suitable question answering subsystems according to their types. Second, the suitable answer summarization strategy can be employed for each question type. This network unified the rule-based and statistical methods into a single framework in fuzzy discriminative learning approach.

Chang et al. in [10] Minimally Supervised Question Classification and Answering based on WordNet and Wikipedia, this method is used for classifying the question into semantic categories in the lexical database like Word Net. In the database, a set of 25 Word Net lexicographer's file is taken from the titles of Wikipedia entry. In this technique surface patterns identification methods is used which classified questions into sets of word-based patterns. Answers are then extracted from retrieved documents using these patterns. Without the help of external knowledge, surface pattern methods suffer from limited ability to exclude answers that are in irrelevant semantic classes, especially when using smaller or heterogeneous corpora. The main focus of this system is to deploy question classification to develop an open domain, general-purpose QA system. Wikipedia titles and Wikipedia categories are used in the process of generating WikiSense.

Zhang et al. [3] considered the machine learning approach for question classification. Support Vector Machine (SVM) replaced the regular expression based classifier with the one that learns from a set of labeled questions. Here, the question represents the *Frequency Weighted Vector* of silent terms. SVM is a binary classifier where the idea is to find a decision surface (or a hyper plane if the training examples are linearly separable) that separates the positive and negative examples while maximizing the minimum margin. The margin is defined as the distance between the decision surface and the nearest positive and negative training examples (called support vectors). The core of SVM is kernel function. Although, SVMs are binary classifiers but they can be extended to solve multi-class classification problems, such as question classification.

A comparative study of available literature shows that classifier proposed have some area of improvements that are stated as follows:

i) The Head Word and SVM in [3,9], classifier has a good generalization performance but in practical point of view it has the high algorithmic complexity and extensive memory requirement in large information retrieval.

ii) The function based classifier in [11], uses the Markov Logic Network which confined the question in a set of six categories like Fact, list, Definition, Reason, Solution and navigation. Hence, it restricts the question into some categories only.

iii) Word Net and Wikipedia based classification in [10], used the redundant lexical database which is itself not very much authenticated.

Above stated drawbacks of different techniques are taken in consideration for classification of user question and indexing of pages for efficient question answering.

## [III]PROPOSED ARCHITECTURE

For efficient question answering, the correct classification of the question is imperative. Then, for answering user's question, it is needed to maintain an index that considers the type of answer user expects from the system. The architecture for Question Classification based indexing schemes for answering user's questions is given in **[Figure-1]** and seven functional modules are discussed as follows:

i) Crawler

ii) Summarizer

iii) Preprocessor

iv) Indexer

v) Question Classifier

vi) Searcher

vii) Ranking

3.1 **Crawler**

The Crawler downloads web pages as well as Blog Pages from Blogosphere. A general web crawler is used for crawling web pages. A Web crawler starts with a list of URLs to visit, called the seeds. As the crawler visits these URLs, it identifies all the hyperlinks in the page and adds them to the list of URLs to visit, called the crawl frontier. URLs from the frontier are recursively visited according to a set of policies. For downloading Blog pages, the crawler used which uses the techniques for crawling blog pages as discussed in [12] are used. This crawler crawls WWW for downloading the Blog Pages as compared to general Web pages that crawl general Web pages. The architecture discussed in [12], uses some of the features which are found to be different for the Blog pages as compared to general Web page. The following are the features:

• The Blog posts are ordered by date, the recent one appears at the top.

• A Blog word is found in URLs of the Blog pages.

• There is RSS tag.

• The most of the hyperlinks point to the Blog itself.

The Crawler for blog pages comprises of four modules given as follows:

i) i) URL Crawler

ii) Blog Checker

iii) Blog Extractor

iv) Link Extractor

The downloaded pages, both general Web pages and Blog pages, are thus stored in page repository, which is then read by the Summarizer module for summarization.

Fig: 1. Proposed Architecture

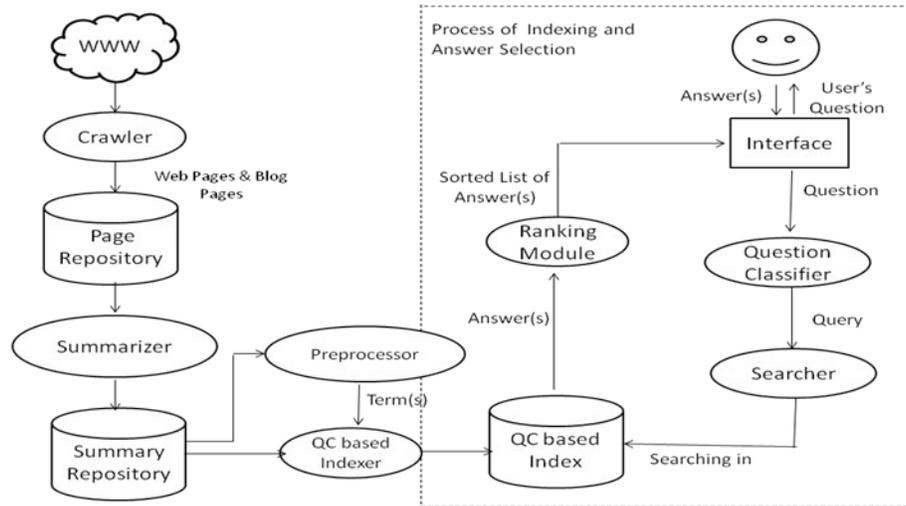

### 3.2 Summarizer

*Summary* is a text segment produced from one or more documents that contain a significant portion of the information in the original text(s), and that is no longer than half of the original text(s). Summary [1] can be created manually or automatically. For automatic summarization, some tools are available online like tool4noobs, text Compactor, Free summarizer etc. **[Table- 1]** describes comparative analysis of summaries generated from different online summarizers.

| Blog URL | http://sordyl.info/2013/01/14/notes-on-distributed-systems-for-young-bloods-something-similar/ |
|---|---|
| **Original Text** | Distributed systems are different because they fail often. Writing robust distributed systems costs more than writing robust single-machine systems. Robust, open source distributed systems are much less common than robust, single-machine systems. Coordination is very hard. Avoid coordinating machines wherever possible. If you can fit your problem in memory, it's probably trivial. "It's slow" is the hardest problem you'll ever debug. Implement backpressure throughout your system. Find ways to be partially available. Metrics are the only way to get your job done. Use percentiles, not averages. Learn to estimate your capacity. Feature flags are how infrastructure is rolled out. Choose id spaces wisely. Exploit data-locality. Writing cached data back to storage is bad. Computers can do more than you think they can. Use the CAP theorem to critique systems. |
| **Text Compactor** | Distributed systems are different because they fail often. Writing robust distributed systems costs more than writing robust single-machine systems. Robust, open source distributed systems are much less common than robust, single-machine systems. If you can fit your problem in memory, it's probably trivial. "It's slow" is the hardest problem you'll ever debug. Implement backpressure throughout your system. Writing cached data back to storage is bad. Use the CAP theorem to critique systems. |
| **Free Summarizer** | Distributed systems are different because they fail often. Writing robust distributed systems costs more than writing robust single-machine systems. Robust, open source distributed systems are much less common than robust, single-machine systems. Use the CAP theorem to critique systems. |
| **Tool4Noobs** | Writing robust distributed systems costs more than writing robust single-machine systems. Robust, open source distributed systems are much less common than robust, single-machine systems. Distributed systems are different |

| | because they fail often. If you can fit your problem in memory, it's probably trivial. "It's slow" is the hardest problem you'll ever debug. Feature flags are how infrastructure is rolled out. Choose id spaces wisely. Exploit data-locality. |
|---|---|

Table: 1. Comparative Analysis of Summaries by online summarizer

Summarization process filters out non-relevant content from the documents. The text produced specifies what the original document is about. The paper user TextCompactor for summarization for Web pages and Blog pages both.

### 3.3 Preprocessor

Summarized documents contain set of sentences that are further processed by Preprocessor module. This module process the summary by using the method of Tokenization, Stop listing and Stemming. Tokenization [13] is the process of parsing the document and extracting tokens. Tokens are the meaningful elements, the words, phrases, symbols. Stoplisting [13] is a process used for removal of stop words and a, an, the etc. which are not meaningful. Stemming [13] is a process of reducing a word into its stem like reducing cars to car, education to educate etc. The function of this module is described in following pseudo code:

*Input: Summary*
*Output: TermSet i.e. set of terms*
**Preprocessor ()**
*{*
*Tokens ← Tokenization(Summary);*
*Stemming (Tokens);*
*TermSet ← Stoplist(Tokens);*
*Return (TermSet);*
*}*

This pseudo code takes summary of Blog pages and general Web pages as its input and then producing set of terms.

### 3.4 Indexer

Indexer takes TermSet generated by the preprocessor as its input and generates the index as shown in **[Table-2]** by using the pseudo code given as follows. The index is based on type of answer expected by the user in response to his question. Index as shown in **[Table-2],** contains Answer Type, Terms, Sid (Sentence Id) and Pid (Page Id).

*Input: TermSet*
*Output: Index*
**Indexer ()**
*{*
*While (! TermSet Empty)*
*{*
*For each Term in TermSet*
*{*
*Generate a query;*
*Define: Term; //by Web Definitions*
*Analyze the definition for Answer type; //Using [Table- 3]*
*Index the Term*
*}*
*}*
*Return(Index)*
*}*

Indexer pseudo code takes each of the term in the TermSet obtained from preprocessor module and generates a query by using the Web Definitions.

When we include "define" in front of a term, Web displays one definition about the term by using online dictionaries and/or WordNet as the source. This Definition provides complete description about the term. Snapshot of the term description given in the Web definition of the term "Continent" is shown in **[Figure-2]**.

Fig: 2 Snapshot of Web Definition of "Continent"

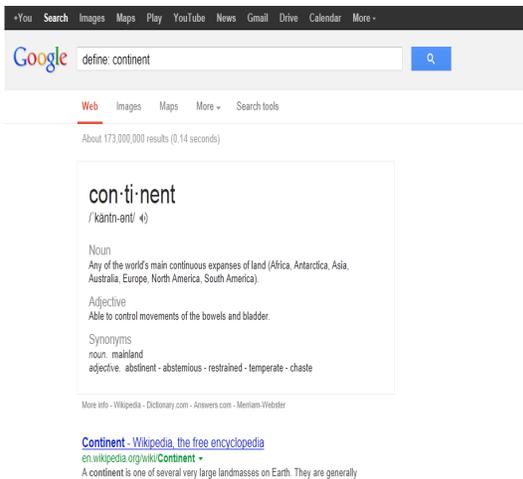

The Web definition is then analyzed to identify the appropriate Answer Type. After analysis, the term description is mapped to the corresponding Answer types using **[Table-3]**. For e.g., if Web definition (shown in **[Figure- 2])** about the tem "*Continent*" is" *Any of the world's main continuous expanses of land (Africa, Antarctica, Asia, Australia, Europe, North America, and South America)*", then since the Web definition describes about *land*, so considering "*land " as the term description, it is* categorized under the "*Location"* answer type.

| Answer Type | Terms | Sid And Pid |
|---|---|---|
| Person | Batsman | s5,p7 |
| Person | President | s1,p9 |
| Location | USA | s4,p7 |
| Procedure | Incoming | s3,p4 |
| Number | Series | s2,p7 |
| Day | Sunday | s1,p7 |
| Month | December | (s4,p7), ( s1,p9) |
| Time | Old | s2,p7 |
| Time | Century | s4,p8 |
| Abbreviation | ADT | (s5,p1), (s1,p2), (s4,p2) |
| Organization | Corporate | s5,p5 |
| Definition | Show | (s2,p9), (s4,p9) |

Table: 2. Question Classification based Index

Then the term is indexed along with the Sentence id and Page id in which it appears under the identified Answer type.

Let us take an example, *Batsman* is indexed under *Person* Answer Type, and it is present in the fifth S*entence* of *summary with id 7.*

| Term description | Answer type |
|---|---|
| Government-Agency, Agency, Company, Airline, University, Institute, Sports-Team | Organization |
| Leader, Father, Mother , Sister, Brother, King, Queen, Emperor, Name | Person |
| City, Country, State, Territory, Mountain, Island, Star, Constellation, Street, Land | Location |
| Currency | Money |

| Full form | Abbreviation |
|---|---|
| Quantity, Distance | Number |
| Procedure, Method, process | Process |
| Day, Days of the week | Day |
| AM, PM | Time |
| Year | Year |
| Months of the year like January etc. | Month |

Table: 3. Terms Description Table

### 3.5 Question Classifier

The Question Classifier [5] module takes the user's question as its input and identifies the question class and performs classification process, discussed as follows:

*Step1. First word of the question identifies the question class.*

*Step2. Remaining part of question is converted into query.*

The proposed architecture restricts the question class as; *Who, What, Where, Where, Which, Why and How*. After step 1, remaining part of the question is converted into query. The process of converting a question into a query is shown in **[Table- 4].**

| Question | Question Class | Query | |
|---|---|---|---|
| | | Answer type | Term set |
| **Who** discovered stem cell | Who | Person, organization | Discover, Stem, Cell |
| **Which** is the coldest place in the world | Which | Person, location, month, time, year, day | Cold, Place, World |
| **When** did titanic sink | When | Time, year, day, month | Titanic, Sink |
| **How** is the president of USA elected | How | Process | President, USA, Elect |
| **Why** did Hitler kill himself | Why | Reason | Hitler, Kill |

Table: 4. Generation of query

**[Table-4]** shows that question classifier divides the question into *Question Class* and *Query*. For e.g. "*Who discovered stem Cell*?", has "*Who*" as question class, "*Person, Organization*" as Answer Type and "*Discover, Stem, Cell*" as TermSet.

### 3.6 Searcher

Searcher takes as its input, question class and query given by the question classifier module and then maps the question class into appropriate Answer type(s) by using **[Table-5].** Then using Question Classification based index given in **[Table-2]**, it searches for the terms in the query corresponding to the found answer type. It then extracts the Sentence id and the Page id in which the terms appear. The sentences are given as answers, which are then passed to the ranking module for assigning a score to each answer. The mapping of Question class into Answer type is given in **[Table-5]**.

35

| Question Class | Answer Type |
|---|---|
| Who | Person, Organization |
| Where | Location |
| What | Money, Number, Definition, Procedure, Abbreviation, Organization, Person, Year, Month, Day, Time, Location |
| When | Time, Year, Day, Month |
| Which | Person, Location, Month, Time, Year, Day |
| Why | Reason |
| How | Process |

Table: 5. Mapping of Question Classification into Answer type

**[Table-5]** describes the Answer type(s) for each question class. The question class is of *Wh-type* or *How*. The answer type(s) corresponding to the question class specifies the focus of the question i.e. what user expects as answer to his question. For e.g. if the user's question is *"Who is the President of USA"*, in this, *"Who"* is Question Class and the corresponding Answer types are Person *and Organization*. Query formed is a set of terms containing P*resident* and *USA*. *Searcher* returns (s1,p9) and (s4,p7) as answers as a result of searching in the index shown in **[Table-2]**.

3.7 **Ranking module**

Set of sentences retrieved from searcher module are the candidate answers to the user's question. These answer(s) are then ranked in appropriate order [6] by the ranking module. The system prompts the user to give a feedback on each answer that appears as a result. Using this approach an appropriate score is assign to each answer returned by the system which is then used for sorting the results. The sorted list of answer then present to the user. The answers which are liked by most of the user are ranked higher and appear at the top of the list.

[IV] EXPERIMENTAL EVALUATION

The proposed work is implemented in Java using Swing. The snapshot of the system has been shown in **[Figure-3]**.

Fig: 3. Snapshot of answer returned by the system

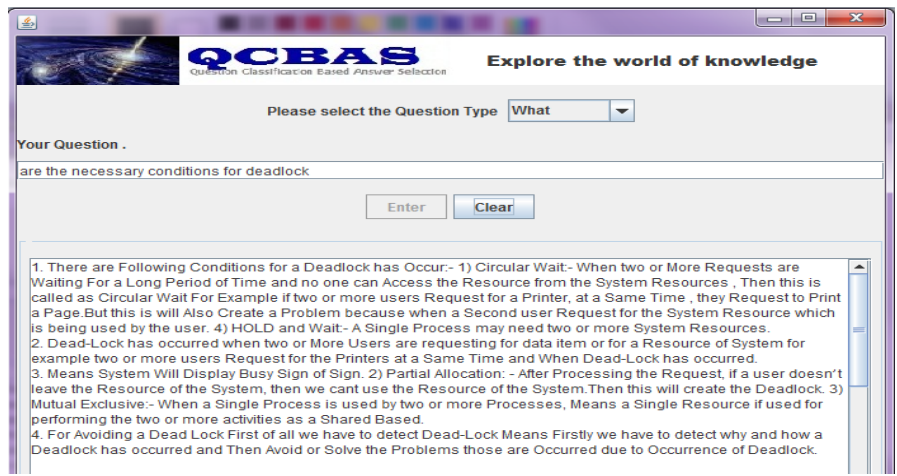

A survey was conducted on Answer Types listed in **[Table-5].** On the basis of the survey, relevant factors for each Answer Types are identified. Then the performance analysis of the system was performed on 270 questions in total where 10 questions were selected belonging to each answer types. The relevance score of each Answer type was calculated by using following formula:

$$\text{Answer Relevance Score} = \frac{RF}{TF} \times 100 \text{ (ARS) (in \%)}$$

Where,

RF: No. of relevant factors returned by the system

TF: Total no. of relevant factors

**[Table-6]** shows no. of relevant factors returned by the system for 10 sample questions of "*Person*" Answer Type and total no. of relevant factors. Also, the relevant score is calculated.

| S.No. | Answer Type | Relevant Factors | Questions | No. of Relevant Factors Returned (RF) | Total No. of Relevant Factors (TF) | ARS = RF/TF * 100 |
|---|---|---|---|---|---|---|
| Q1 | Person | 1. Person's Name 2. Education 3. Birth Place/ native Place 4. When he/she was born/died 5. His/her contribution 6. Other related information | Who is the first woman Prime Minister of India? | 5 | 6 | 83.33 |
| Q2 | Person | | Who is the founder of Infosys? | 4 | 6 | 66.67 |
| Q3 | Person | | Who is Mother Teresa? | 6 | 6 | 100.00 |
| Q4 | Person | | Who is Barak Obama? | 5 | 6 | 83.33 |
| Q5 | Person | | Who is Michael Jackson? | 3 | 6 | 50.00 |
| Q6 | Person | | Who is the inventor of Telephone? | 4 | 6 | 66.67 |
| Q7 | Person | | Who is the second man on the Moon? | 4 | 6 | 66.67 |
| Q8 | Person | | Who discovered America? | 5 | 6 | 83.33 |
| Q9 | Person | | Who is known as the "Missile Man"? | 5 | 6 | 83.33 |
| Q10 | Person | | Who is the first mughal emperor of India? | 6 | 6 | 100.00 |

Fig: 4. Graph sharing Answer Relevance Score for sample Question for "*Person*" Answer Type

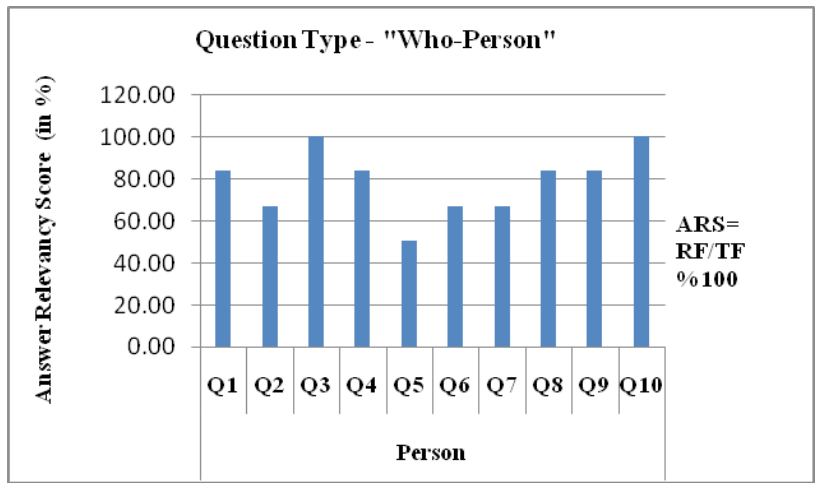

| S.No. | Question Class | Answer Type | ARS (in %) RF/TF*100 |
|---|---|---|---|
| 1 | Who | Person | 78.33 |
| | | Organization | 79.00 |
| 2 | Where | Location | 81.27 |
| 3 | When | Time | 76.45 |
| | | Year | 79.00 |
| | | Day | 85.00 |
| | | Month | 83.00 |
| 4 | What | Money | 84.2 |
| | | Number | 81.00 |
| | | Definition | 84.00 |
| | | Procedure | 83.00 |
| | | Abbreviation | 80.10 |
| | | Organization | 80.00 |
| | | Person | 83.66 |
| | | Year | 82.00 |
| | | Month | 83.00 |
| | | Day | 79.33 |
| | | Time | 83.76 |
| | | Location | 79.00 |
| 5 | Which | Person | 84.00 |
| | | Location | 77.00 |
| | | Month | 83.00 |
| | | Time | 77.00 |
| | | Year | 80.00 |
| | | Day | 84.00 |
| 6 | Why | Reason | 76.88 |

| 7 | How | Process | 80.12 |

Table: 7. Relevancy of different Answer Types

Average Answer Relevance Score for each Answer Type as listed in **[Table-7]** is shown with the help of graphs (shown in **[Figure 5 to 10]**).

Fig: 5. Graph showing Average ARS of "Peron" and "Organization"

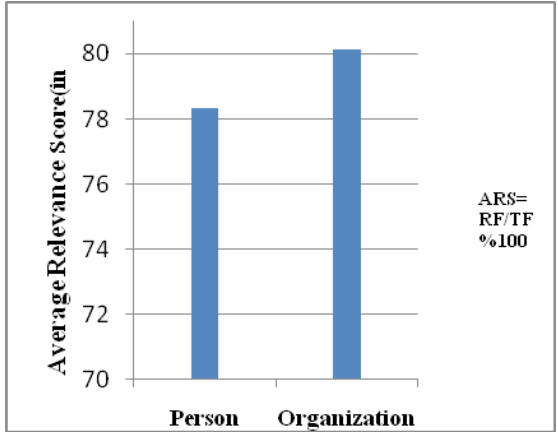

Fig: 6. Graph showing Average ARS of "Location"

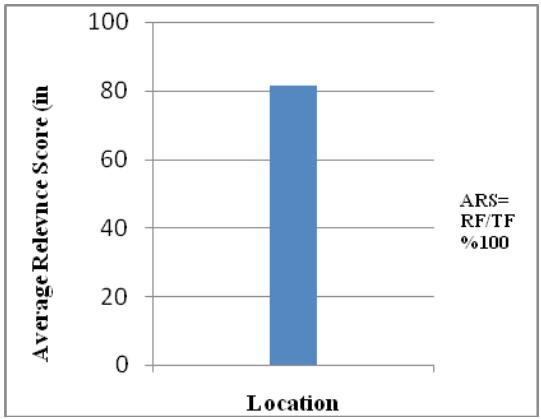

Fig: 7. Graph showing Average ARS of "When" Question Type

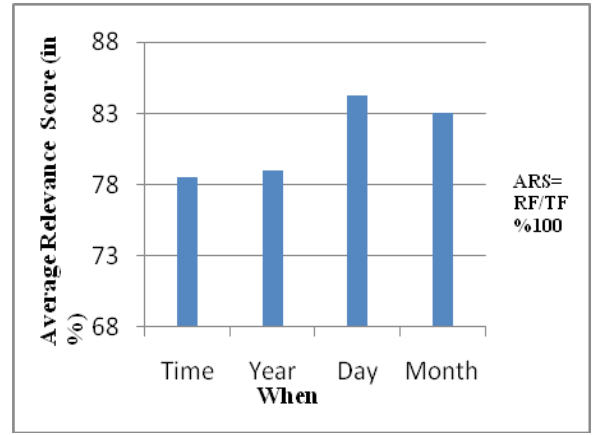

Fig: 8. Graph showing Average ARS of "What" Question Type

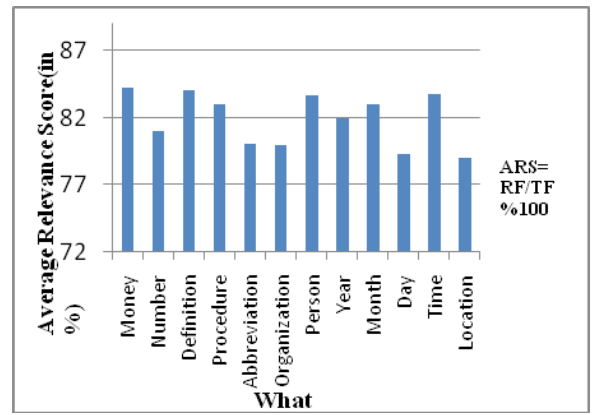

Fig: 9. Graph showing Average ARS of "Which" Question Type

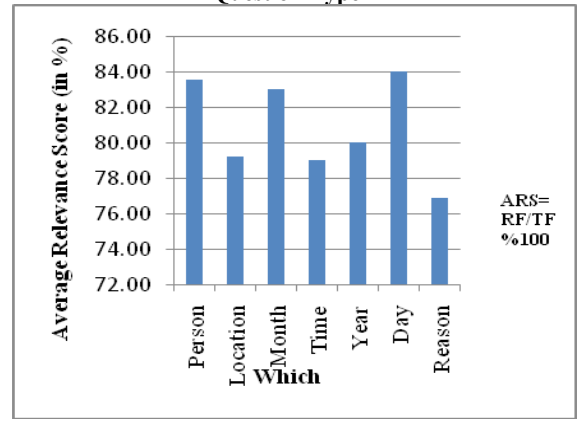

Fig: 10. Graph showing Average ARS of "Why" and "How" Question Types

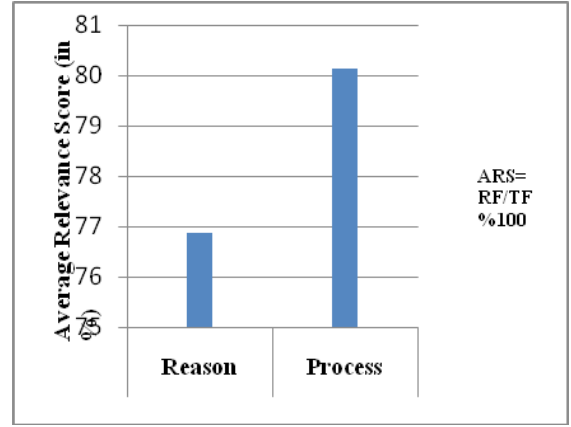

On analysis, it was observed that Average Answer Relevance Score (ARS) was found in the range from 75.98% to 85.77%. The experimental results from the system are found to be promising and the shows better performance than the existing systems.

## [V] DISCUSSIONS – A COMPARATIVE REVIEW

The experimental evaluation of the system shows the performance of the system as a whole. There are other question answering systems present like QUALLIFIER, TextMap, QuASM and START. We compared our system with START QA system which is a natural language based Answer selection system. The system is compared by asking a set of 10 questions, and the answers are analyzed by taking relevant factors obtained from the survey, and the results shows promising behaviors of this system as compared to START. Table 8 illustrates the records for the average relevance score for QCBAS and START. And Fig. 11 shows the Average Relevance score question classification based Answer Selection (QCBAS) system and START.

| Question's S.No. | Total No. Relevant Factors | No. of Relevant factors (START) | No. of Relevant factors (QCBAS) | ARS for START | ARS for QCBAS |
|---|---|---|---|---|---|
| Q1 | 7 | 5 | 6 | 71.43 | 85.71 |
| Q2 | 7 | 4 | 7 | 57.14 | 100.00 |
| Q3 | 7 | 5 | 6 | 71.43 | 85.71 |
| Q4 | 7 | 6 | 5 | 85.71 | 71.43 |
| Q5 | 2 | 1 | 2 | 50.00 | 100.00 |
| Q6 | 2 | 1 | 2 | 50.00 | 100.00 |
| Q7 | 7 | 4 | 6 | 57.14 | 85.71 |
| Q8 | 7 | 5 | 7 | 71.43 | 100.00 |
| Q9 | 7 | 4 | 6 | 57.14 | 85.71 |
| Q10 | 8 | 6 | 8 | 75.00 | 100.00 |

Table 8. Relevancy Answer Score for START and QCBAS

Above table shows that 10 question are taken and a comparative study on these systems is performed. In the process of comparison, relevant factors for different answers types are taken from the survey. The relevant factors are further used for deciding the average relevance factor (ARS). The corresponding ARS is calculated and on the basis of this table, a bar chart is drawn. The performance graph is shown in the Fig. 11.

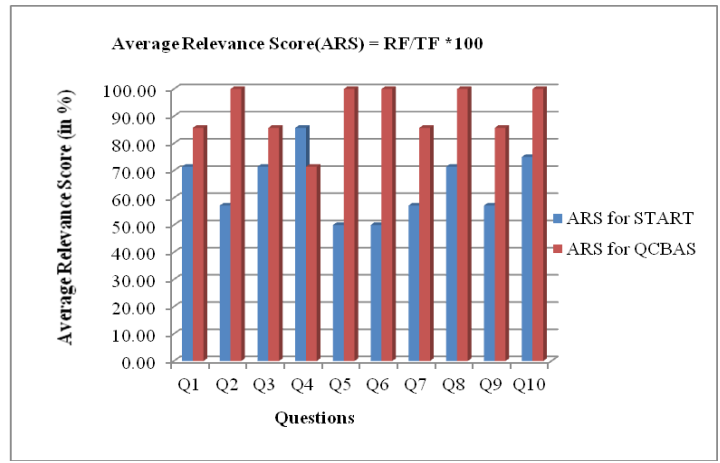

**Fig. 11. Graph showing Comparison of START and QCBAS System**

This figure describes the performance by plotting 3-D clustered column bar chart of these two systems. The red column shows the performance evaluation of our system which is quite appreciable. The average rate of correct answer given by START is 62.15% and the average rate of QCBAS is 79.17%, which is good. Hence, these experimental results shows that this system performance better than the existing systems like START.

**[VI] CONCLUSION**

Question classification being a crucial part of question answering system classifies the question into an appropriate question class and then maps it into expected answer type. This approach is used to extract the answer(s) for the user's question on the basis of expected answer type. The paper proposes a novel architecture for Question Answering and pseudo code for Question Classification based indexing. The technique discussed is different from traditional classification techniques as it is based on indexing the documents on the basis of expected answer type identified by using the Web Definitions. The experimental results show that the answers returned by the system have higher relevance score. Thus, the system shows better performance as compared to existing question answering systems.